\documentclass[sigconf,natbib=true]{acmart}

\AtBeginDocument{%
  \providecommand\BibTeX{{%
    \normalfont B\kern-0.5em{\scshape i\kern-0.25em b}\kern-0.8em\TeX}}}

\setcopyright{acmcopyright}

\copyrightyear{2022} 
\acmYear{2022} 
\setcopyright{acmcopyright}\acmConference[CIKM '22]{Proceedings of the 31st ACM International Conference on Information and Knowledge Management}{October 17--21, 2022}{Atlanta, GA, USA}
\acmBooktitle{Proceedings of the 31st ACM International Conference on Information and Knowledge Management (CIKM '22), October 17--21, 2022, Atlanta, GA, USA}
\acmPrice{15.00}
\acmDOI{10.1145/3511808.3557351}
\acmISBN{978-1-4503-9236-5/22/10}

\usepackage{enumitem}
\usepackage{tabularx}
\usepackage{makecell}
\begin{document}

\title{
Hierarchically Fusing Long and Short-Term User Interests for Click-Through Rate Prediction in Product Search
}






\author{Qijie Shen}
\affiliation{%
  \institution{Alibaba Group}
  \city{Hangzhou}
  \country{China}
}
\email{qjshenxdu@gmail.com}

\author{Hong Wen}
\affiliation{%
  \institution{Alibaba Group}
  \city{Hangzhou}
  \country{China}
}
\email{qinggan.wh@alibaba-inc.com}

\author{Jing Zhang}
\affiliation{%
  \institution{The University of Sydney}
  \city{Darlington NSW 2008}
  \country{Australia}
}
\email{jing.zhang1@sydney.edu.au}

\author{Qi Rao}
\affiliation{%
  \institution{Alibaba Group}
  \city{Hangzhou}
  \country{China}
}
\email{rq145781@alibaba-inc.com}

\def\authors{Qijie Shen, Hong Wen, Jing Zhang, Qi Rao}
\renewcommand{\shortauthors}{Qijie Shen, Hong Wen, Jing Zhang, \& Qi Rao.}
\begin{abstract}

Estimating Click-Through Rate (CTR) is a vital yet challenging task in personalized product search. However, existing CTR methods still struggle in the product search settings due to the following three challenges including how to more effectively extract users' short-term interests with respect to multiple aspects, how to extract and fuse users' long-term interest with short-term interests, how to address the entangling characteristic of long and short-term interests. To resolve these challenges, in this paper, we propose a new approach named Hierarchical Interests Fusing Network (HIFN), which consists of four basic modules namely Short-term Interests Extractor (SIE), Long-term Interests Extractor (LIE), Interests Fusion Module (IFM) and Interests Disentanglement Module (IDM). Specifically, SIE is proposed to extract user's short-term interests by integrating three fundamental interests encoders within it namely query-dependent, target-dependent and causal-dependent interest encoder, respectively, followed by delivering the resultant representation to the module LIE, where it can effectively capture user long-term interests by devising an attention mechanism with respect to the short-term interests from SIE module. In IFM, the achieved long and short-term interests are further fused in an adaptive manner, followed by concatenating it with original raw context features for the final prediction result. Last but not least, considering the entangling characteristic of long and short-term interests, IDM further devises a self-supervised framework to disentangle long and short-term interests. Extensive offline and online evaluations on a real-world e-commerce platform demonstrate the superiority of HIFN over state-of-the-art methods.

\end{abstract}




\maketitle

\section{Introduction}

With the explosive emergence of products in e-commerce platforms, product search serves an indispensable role in discovering desirable items that satisfy users. When a user submits a query, the product search will deliver a small yet well ranked product set from billions of candidates to end-users through a typical multi-stage pipeline of ``match$\to$prerank$\to$rank$\to$rerank'' \cite{li2021embedding,li2020hierarchical,zhang2020empowering}. In this paper, we only focus on the Click Through-Rate (CTR) prediction task in the \emph{rank} stage, which aims to predict the probability of users clicking items and has a great impact on improving user experience and boosting the revenue of e-commerce platforms. 

The prerequisite for a CTR prediction task in personalized product search is to accurately extract the user preferences from historical behaviors and effectively integrate them with the current query \cite{guo2019attentive}. Besides, it has been well recognized that there are two types of user preferences \cite{bennett2012modeling}, $i.e.$, long and short-term ones, where the former exhibits users' inherent and relatively stable (evolving slowly) preferences, such as preferred color, fitting size, price preferences, imperceptibly influenced by the user's family background, age, marital status, education, $etc$. While the short-term interests convey user preference intention in a relatively short period, which can be inferred from their recent behaviors and also affected by incidentally transient events, such as new product release, season change and special personal occasions like birthday \cite{xiang2010temporal}. In a nutshell, users' short-term interests evolve in a more high-frequently and drastically manner compared with the long-term ones.

A typical paradigm of existing personalized product search models is to represent the user intents and items with embedding vectors explicitly, followed by matching them in the latent space with dot product or feeding them to neural layers to yield the prediction score \cite{bi2020transformer}. For example, HEM \cite{ai2017learning} fuses user vector and query vector to represent user intent with a convex combination. AEM \cite{ai2019zero} is the first attention-based embedding model which constructs query-dependent user embeddings by employing attention mechanism to users’ historical purchased items with respect to the query. ZAM \cite{ai2019zero} devises a zero attention mechanism to determine when and how to conduct personalization under various scenarios. TEM \cite{bi2020transformer} dynamically control the influence of personalization by encoding the sequence of query and user’s historical behaviors with a transformer architecture. Guo \cite{guo2019attentive} proposes a novel Attentive Long Short-Term Preference model (abbreviated as ALSTP) to learn and integrate the long and short-term user preferences with respect to the current query in a typical personalized product search scenario. Despite its effectiveness, we argue that the existing product search paradigm has several obvious limitations.

First, existing methods always extract users' interest representation by aggregating user historical behaviors with respect to the query submitted by users. We argue that the strategy can work efficiently only when users exhibit their demands with submitted queries explicitly. For example, when a user submits a query ''red canvas shoes'', it can express the user's requirements on color, type, and material clearly, resulting in the effectiveness of extracting query-dependent interest by straightforwardly employing attention mechanism to user behaviors with respect to the submitted query. However, users sometimes could express their demands with obscure intention. For example, when a user issues a query ''gift'', which indicates the user does not specify the brand, category, style, price, $etc.$ he (she) is desiring at this moment, resulting in the various types of candidate items delivered by previous stage \emph{match}. Therefore, we need to extract user's current precious interest delivered from historical behaviors with respect to individual target item. In other words, an alternative solution is to employ attention mechanism with respect to target product instead of query in such situation to extract target-dependent user interest. 

Second, there exists various types of historical behaviors, such as \emph{click}, \emph{favorite}, or \emph{purchase}, which are inadvertently neglected by previous works or just regarded as certain a kind of behavior features integrated into original framework. We argue multiple kinds of behaviors contain abundant information that is worth investigating deeply for superior user interests representation. However, it is not trivial to integrate them into original framework. For example, the purchase behaviors delivered by some users denotes their recent desired demands have been satisfied eventually, which implies a smaller weight on the purchase behavior contributing to the final representation of user interests. On the contrary, some users probably click, or favorite after a certain click behavior on a special category, which denotes the intense of users' interests are continuously increasing, resulting in a higher weight should be given to this kind of behavior for final representation of user interest. In this paper, we regard the phenomenon, $i.e.$, the causality of multiple types of user behaviors, as causal-dependent user interest.


Third, although the state-of-the-art long and short- term interests method, $i.e.$, ALSTP, has achieved a significant performance, it still struggle in achieving better performance since it only considers the query-dependent user interest, neglecting the target-dependent interest and the causal-dependent interest. Therefore, it will lead to inferior performance in some cases. Besides, ALSTP just concatenates the long and short-term interests and entangles both aspects together for final prediction without explicitly distinguishing the importance of long and short-term interests to final prediction, which will lead to inferior accuracy and interpretability.


To address aforementioned challenges, we proposed a novel method named Hierarchical Interests Fusing Network (HIFN) for CTR prediction in product search, which consists of four basic modules namely Short-term Interests Extractor (SIE), Long-term Interests Extractor (LIE), Interests Fusion Module (IFM) and Interests Disentanglement Module (IDM). Specifically, SIE is proposed to extract user's short-term interests by employing three interest encoders, $i.e.$, query-dependent interest encoder (QDIE), target-dependent interest encoder (TDIE) and causal-dependent interest encoder (CDIE) to extract query-dependent, target-dependent and causal-dependent interests, respectively, followed by fusing these three interests in an adaptive manner. In LIE, we capture user long-term interest by devising an attention mechanism with respect to the output representation of SIE module, followed by fusing the long and short-term interest in module IFM. Next, all the representation features are concatenated with original raw context features and fed into fully connected layers to generate the final prediction results. Last but not least, motivated by DLSR \cite{zheng2022disentangling}, IDM injects pseudo labels for both long and short-term interest and employs an self-supervised framework to disentangle long and short-term interests. But different from DLSR, we devise an updating strategy to set long-term interest proxy, which is more suitable and achieves better performance than the settings in DLSR.The contribution of this paper is three-fold:


\begin{itemize}[leftmargin=*]
    \item We propose a novel method named HIFN for CTR prediction in product search, which facilitates learning long and short-term user interests representations in a hierarchical manner.

    \item We devise four key modules including SIE, LIE, IFM and IDM to jointly address the aforementioned challenges in the context of product search, which can be implemented efficiently in a multi-task end-to-end learning framework.
    
    \item We conduct extensive experiments on real-world offline datasets and online A/B test. Both results demonstrate the effectiveness of HIFN over state-of-the-art methods. 

\end{itemize}

\begin{figure*}
  \centering
  \includegraphics[width=\textwidth]{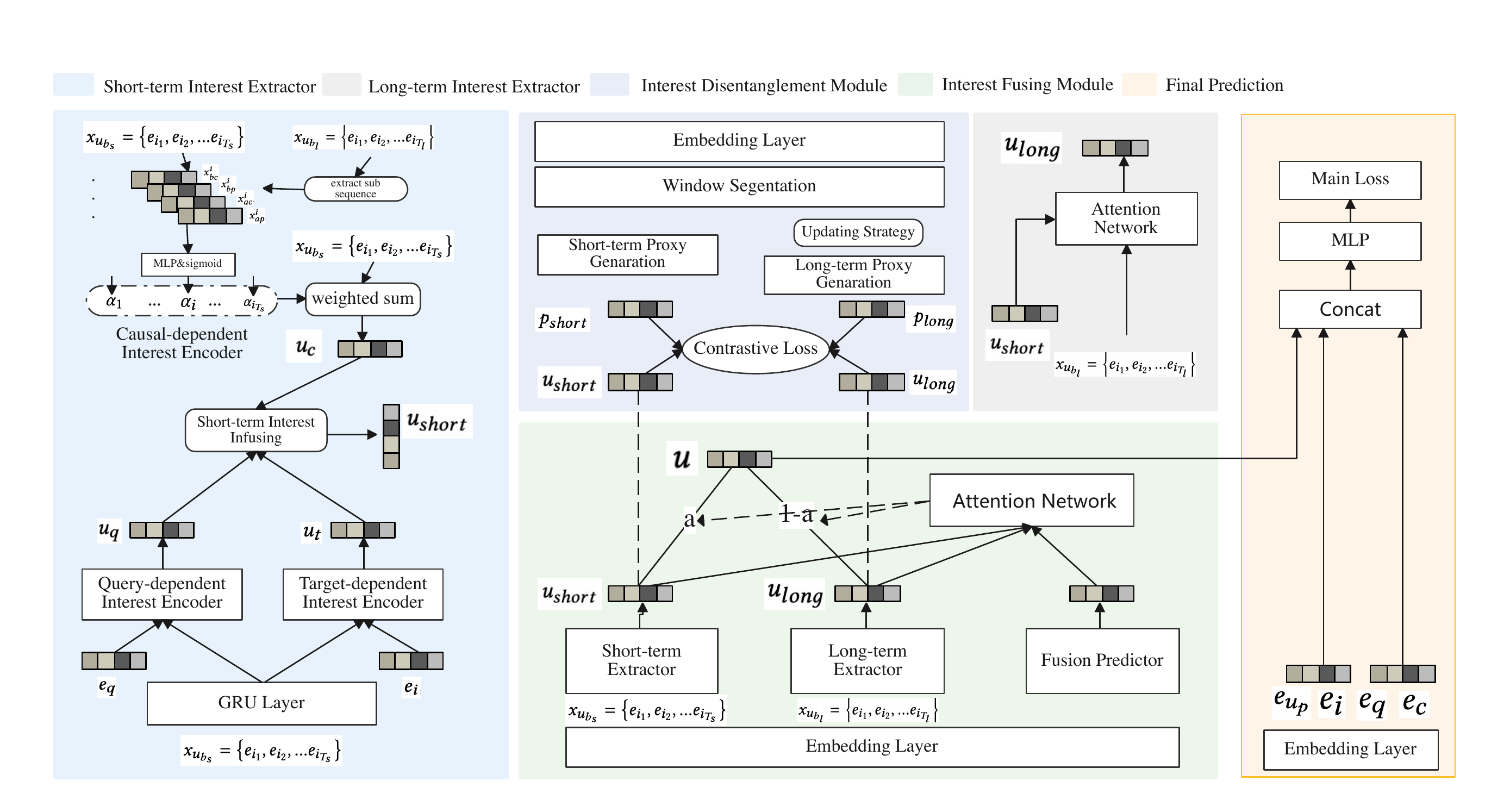}
  \caption{The overview architecture of the proposed HIFN. It consists of several modules, i.e., SIE, LIE, IFM and IDM.
  }
  \label{fig:model_pic}
\end{figure*}

\section{Proposed Method}

In this paper, we propose a novel method named HIFN, as 
depicted in Figure~\ref{fig:model_pic}. It consists of four basic modules namely Short-term Interest Extractor (SIE), Long-term
Interest Extractor (LIE), Interest Fusion Module (IFM) and Interest Disentanglement Module (IDM). We will detail them as follows.

\subsection{Problem Definition}
\label{subsec:problem_definition}

Given user profiles $u_{p}$, long-term user behaviors $u_{b_{l}}$, short-term user behaviors $u_{b_{s}}$, target item $i$, search query $q$, context feature $c$, where $u_{b_{l}}=\left \{ i_{1},i_{2},...i_{T_{l}} \right \}$, $u_{b_{s}}=\left \{ i_{1},i_{2},...i_{T_{s}} \right \}$, and $i_{t}$ from set $u_{b_{l}}$ (resp. $u_{b_{s}}$) represents user's $t$-th interacted item from long-term (resp. short-term) behaviors. Besides, $T_{l}$ (resp. $T_{s}$) denotes the length of user's long-term (resp. short-term) behaviors. Next, we firstly encode them into one-hot high-dimensional vectors, followed by transforming these vectors into low-dimensional dense vectors by ways of multiplying corresponding embedding matrices \cite{tao2022sminet,tao2022online,xu2022odnet}. After concatenating various types of dense vectors, we obtain the embedding vectors $e_{u_{p}}$, $x_{u_{b_{l}}}$, $x_{u_{b_{s}}}$, $e_{i}$, $e_{q}$, $e_{c}$ for $u_{p}$, $u_{b_{l}}$, $u_{b_{s}}$, $i$, $q$, $c$, respectively, where $x_{u_{b_{s}}}=\left \{ e_{i_{1}},e_{i_{2}},...e_{i_{T_{s}}} \right \}$, $x_{u_{b_{l}}}=\left \{ e_{i_{1}},e_{i_{2}},...e_{i_{T_{l}}} \right \}$, and $e_{i_{t}}$ is the embedding vector of user's $t$-th interacted item. Our goal is to devise a model to predict the probability of user $u$ clicking target item $i$ given search query $c$ and corresponding relevant features, formally defined as: $\hat{y}_{u,i}=\mathcal{F}(e_{u_{p}},x_{u_{b_{l}}},x_{u_{b_{s}}},e_{i},e_{q},e_{c};\theta ) $, where $\mathcal{F}$ is the learning objective with model parameters $\theta$.


\subsection{Short-Term Interest Extractor}

In general, users' short-term behaviors play a vital role to infer their recent interests. Although existing works towards extracting query-dependent interest in product search have showed significant performance, they still struggle in explicitly delivering users' interests due to the confusion of users' multiple interests in some conditions. Therefore, besides query-dependent interest extraction, an additional supplement is to extract corresponding interest representation from users' behaviors with respect to individual candidate product, called \emph{target-dependent} interest extraction. In addition, users always exhibit various types of behaviors, $i.e.$, \emph{click} or \emph{purchase}, while neglected by previous works when extracting users' interests from their behaviors or just regarded as certain a kind of behavior features. And we argue users' multiple types of behaviors are very essential for users' final interest extraction, called \emph{causal-dependent} interest extraction.


Based on the above considerations, we first adopt a GRU \cite{cho2014properties} to model the short-term interests, as it has been successfully applied in the session-based recommendation \cite{hidasi2015session}. The RNN module takes the recent $s$ product representations $x_{u_{b_{s}}}=\left \{ e_{i_{1}},e_{i_{2}},...e_{i_{T_{s}}} \right \}$ as inputs. We thus
can obtain a set of high dimensional hidden representations $x_{u_{b_{s}}}^{,}=\left \{ e_{i_{1}}^{,},e_{i_{2}}^{,},...e_{i_{T_{s}}^{,}} \right \}$, denoting the short-term user preference.

Then, We feed $x_{u_{b_{s}}}^{,}$ into three well-designed interest encoder named Query-dependent Interest Encoder, Target-dependent Interest Encoder, Causal-dependent Interest Encoder, respectively. 

\subsubsection{Query-dependent Interest Encoder}\label{sec:qdie}

Given users express their intention with submitted queries explicitly, we can straightforwardly devise a module named \emph{Query-Dependent Interest Encoder} (QDIE) to accurately extract users' current interests from their behaviors with respect to the submitted query. After considering the varied relevance of historical behaviors with respect to the current submitted queries, we resort to the attention mechanism in QDIE module to extract query-dependent interest, which can be formulated as:

\begin{equation} \label{eq1}
\begin{split}
u_{q} = \sum_{i\in x_{u_{b_{s}}}'}{\frac{exp(f(e_{q},e_{i}))}{\sum_{i'\in x_{u_{b_{s}}}'}exp(f(e_{q},e_{i'}))}{}}e_{i}
\end{split}
\end{equation}
where $f(q, i)$ is an attention function that defines the
weight of each item $i$ with respect to the current query $q$, defined as:
\begin{equation} \label{eq2}
\begin{split}
f(q,i)=(i\cdot tanh(W_{f}\cdot q+b_{f}))\cdot W_{h}
\end{split}
\end{equation}
where $W_{h}\in\mathbb{R}^{\beta}$, $W_{f}\in\mathbb{R}^{\alpha\times \beta\times\alpha}$, $b_{f}\in\mathbb{R}^{\alpha\times\beta}$.


\subsubsection{Target-dependent Interest Encoder}

Practically, users sometimes could not clearly express their search intention with submitted queries. Under such circumstances, we employ a module named \emph{Target-Dependent Interest Encoder} (TDIE) to extract users' current interest from their behaviors with respect to individual target item. Take the travel scenario as example, when a user has many online clicking behaviors on the scenic spots from city \emph{Hangzhou} and after he(she) submitted a query named \emph{Hangzhou}, we can probably infer the user prefers the well-known spot \emph{West Lake} more compared with \emph{snack} in \emph{Hangzhou}. Similar with \nameref{sec:qdie}, we directly employ an attention mechanism on $x_{u_{b_{s}}}^{'}$ with respect to target item $e_{i}$ to achieve target-dependent interest $u_{t}$.


\subsubsection{Causal-dependent Interest Encoder}

Users usually exhibit diverse and continuous interests by ways of various types of behaviors, $i.e.$, \emph{click} or \emph{purchase}. For example, if users are impressed by \emph{clothes}, they will naturally deliver continuous behaviors on \emph{clothes} related products for a period of time until their needs are satisfied, $i.e.$, desired clothes have been bought eventually. Mathematically, let $e_{i}^{t}$ in user behaviors $x_{u_{b_{s}}}$ denote the category of the $i$-th behavior item at time $t$. In fact, we can extract four auxiliary sub-sequences from $x_{u_{b_{l}}}$ for the behavior item at time $t$ according to whether click (or purchase) behaviors before(or after) time $t$ have the same category with $e_{i}^{t}$, $i.e.$, click sub-sequence before time $t$, $x_{bc}^{i}$, purchase sub-sequence before time $t$, $x_{bp}^{i}$, click sub-sequence after time $t$, $x_{ac}^{i}$, purchase sub-sequence after time $t$, $x_{ap}^{i}$. We argue that some users would express intense interests since they continuously exhibit click behaviors with the same category to the one at $t-$th behavior, whether before or after $t$, especially when users have finished purchase behavior at time $t$. Meanwhile, some users may have continuous click behaviors before time $t$ with the same category to the one at time $t$, while they show the decrement of interests after they finished purchase behaviors. Therefore, we employ the four auxiliary sub-sequences to portray users' precious intention at time $t$, formulated as:

\begin{equation}
\begin{split}
    \alpha_{i} = Sigmoid(FFN( Sum(x_{bc}^{i})||Sum(x_{bp}^{i})||Sum(x_{ac}^{i})||Sum(x_{ap}^{i})))
\label{eq:lstm_form5}
\end{split}
\end{equation}

where $Sum( \cdot )$ indicates sum pooling operator, the reason for using sum pooling is that it can express the density of behavior and achieving great performance, however the computational complexity is lower than others (e.g. GRU, LSTM).  
  Then, causal-dependent Interest is calculated as follows:
  
\begin{equation}
\textit u_{c} = \sum_{i}^{}\alpha _{i}e_{i},
\end{equation}




\subsubsection{Short-term Interest Fusing}
As mentioned earlier, the three aspects both play an important role under different user intentions, we employ a gate network to fusing three aspects of short-term interest. The structure of the gate network is based on a single-layer feed-forward network with a SoftMax activation function. It acts as a selector to calculate the weighted sum of the selected vector. 

\begin{equation}
u_{short} = SoftMax(FFN(u_{t},u_{q},u_{c}))^{T}[u_{t},u_{q},u_{c}],
\end{equation}

\subsection{Long-Term Interests Extractor}

Generally, users' behaviors are categorized into two disjoint groups, $i.e.$, long-term behaviors and short-term ones. When users' short-term behaviors are sparse or even absent, the long-term behaviors will be as a kind of supplement cues to discover their interests. Moreover, users' long-term behaviors are regarded as the reflection of their intrinsic tastes while evolving slowly over time, which further affects their short-term behaviors in a relatively recent period. Motivated by this, we propose a Long-Term Interest Extractor (LTIE) module to extract users' long interests. Specially, we employ an attention mechanism to long-term behaviors with respect to the generated short-term interest $u_{short}$ like Eq. ( \ref{eq1} ) and Eq. ( \ref{eq2} ) to deliver users' long-term interests $u_{long}$.


\subsection{Interests Fusing Module}

Given users' long- and short-term interests, how to fuse them is not trivial due to the different contributions with respect to individual users. One straightforward solution is employing concatenation operator to integrate them, while it still struggles in how to effectively distinguish the contribution of each interest. For example, basketball fans may continuously click basketball shoes when a query "shoes" submitted due to their long-term interest even they have delivered a lots of canvas shoes related behaviors in a recent period. Alternatively, we devise a module named Interest Fusing Module to seamlessly integrate both interest representation together with submitted query, historical behaviors, target item. Formally, the process can be formulated as :


\begin{equation}
\alpha=sigmoid(FFN(e_{i}||e_{q}||u_{short}||u_{long}||\tau_{u_{b_{l}}}))
\end{equation}

\begin{equation}
\tau _{u_{b_{l}}}=GRU(x_{u_{b_{l}}})
\end{equation}

where, $\alpha$ denotes the estimated fusion weight based on aforementioned information. Then,
the final user interest representation can be formulated as follows.

\begin{equation}
u = \alpha\cdot u_{short} + (1-\alpha)\cdot u_{long}
\end{equation}

Finally, we concatenate $u$, $e_{i}$,  $e_{q}$ and $e_{c}$ to form a comprehensive vector, followed by feeding it to a MLP layer with \emph{Relu} as activation function to get the final predicted CTR score $\hat{y}$. Finally, we define the objective function as follows:
\begin{equation}
    Loss_{t}=-\frac{1}{N} \sum_{(x,y)\in S}(ylog\hat{y}+(1-y)log(1-\hat{y})),
\label{eq:din_loss}
\end{equation}
where $S$ denotes the training set with total size $N$, and $y\in \left \{ 0,1 \right \} $ is the ground truth label representing whether click or not.

\subsection{Interests Disentanglement Module}
Desired by Zheng \cite{zheng2022disentangling} who introduces a self-supervised framework for disentanglement, we adapt it for product search. We take the mean representation of short-term behaviors as the proxy of short-term interests. As for long-term interests, we argue that directly mean representation of entire long-term behaviors applied by Zheng \cite{zheng2022disentangling} will lead to sub-optimal performance. Long-term user preference is relatively stable and updates slowly, if we mean representations simply, the temporal variability of long-term interests would be ignored. Since short-term interests are based on the most recent $s$ interacted items, the impact of this drawback is minimal. We apply an updating mechanism to the extraction of long-term interests proxy to capture the slowly evolving characteristic of long-term interests. For simplicity, we use the first $Ts$ products to
set it up, and then update it with every $Ts$ product interacted (evolving gradually). In this setting, $p$ is updated based on a session of $Ts$ products, which is also the window size of modeling the short-term user preference. It is illustrated as follows:
\begin{equation}
   p_{long} = (1-\beta)p_{long} + \beta FFN(h_{t-1})
\label{eq:din_loss}
\end{equation}
where, FFN is multiple layers of feed forward network, $h_{t-1}$ is the mean representation of previous $Ts$ interacted items. As for short-term proxy, it is calculated as follows:

\begin{equation}
   p_{short} = MeanPooling\left \{ e_{i_{1}},e_{i_{2}},...e_{i_{T_{s}}} \right \}
\label{eq:din_loss}
\end{equation}

With proxies serving as labels, we can utilize them to supervise the disentanglement of long and short-term interests. Specifically, we perform contrastive learning \cite{liu2021anomaly} between the outputs of interest extractors and proxies,which requires the learned representations of long and short-term interests to be more similar to their corresponding proxies than the opposite proxies. Formally, there are four contrastive tasks as follows,

\begin{equation}
  sim(u_{long},p_{long}) > sim(u_{long},p_{short})
\label{eq:sim1}
\end{equation}
\begin{equation}
  sim(p_{long},u_{long}) > sim(p_{long},u_{short})
\label{eq:sim2}
\end{equation}
\begin{equation}
  sim(u_{short},p_{short}) > sim(u_{short},p_{long})
\label{eq:sim3}
\end{equation}
\begin{equation}
  sim(p_{short},u_{short}) > sim(p_{short},u_{long})
\label{eq:sim4}
\end{equation}

where Eqn (\ref{eq:sim1})-(\ref{eq:sim2}) supervise long-term interests, and Eqn (\ref{eq:sim3})-(\ref{eq:sim4}) supervise short-term interests, and $sim(\cdot, \cdot)$ measures embedding similarity. we add self-supervision on long and short-term interests modeling which can achieve stronger disentanglement
compared with existing unsupervised approaches. We implement a pairwise loss functions based on Bayesian Personalized Ranking (BPR) to accomplish contrastive learning, are computed as follows:

\begin{equation}
  \text{\pounds}_{bpr}(a,p,q) = softplus(<a,q> - <a,p>)
\label{eq:sim5}
\end{equation}

\begin{equation}
\begin{split}
  \text{\pounds}_{con} = \text{\pounds}_{bpr}(u_{long},p_{long},p_{short}) +  \text{\pounds}_{bpr}(p_{long},u_{long},u_{short}) + \\
  \text{\pounds}_{bpr}(u_{short},p_{short},p_{long}) + 
  \text{\pounds}_{bpr}(p_{short},u_{short},u_{long}).
\label{eq:sim6}
\end{split}
\end{equation}

We train the model in
an end-to-end manner with multi-task learning on two objectives.
Specifically, the joint loss function with a hyper-parameter $\lambda$ to
balance objectives, formulated as follows:
\begin{equation}
  \text{\pounds} = Loss_{t} + \lambda\text{\pounds}_{con}
\label{eq:sim5}
\end{equation}

\section{Experiment}
To verify the effectiveness of HIFN, we conduct extensive experiments to compare it with representative SOTA methods on both offline datasets and online deployment. 

\subsection{Dataset}
We use two datasets to conduct experiments, including a public
 dataset and an industrial dataset namely Amazon and Fliggy, respectively, which are both with million users scale and collected from real-world applications. The details of the adopted datasets are introduced as follows. Statistics of datasets are listed in Table \ref{stat}.

\textbf{Fliggy\footnote{One of the most popular online travel platforms in china, www.fliggy.com}}
The Fliggy dataset which contains properties of users, queries, user historically behaviors ( including click, favorite, and purchase ), and target product, is generated based on user logs collected from October 1th to October 31th, 2021. Note that users with fewer than 20 historical interactions are removed. User logs of the first 30 days in October 2021 is used as the training data, while reserve
the last day in October for validation (before 12 pm) and test (after 12 pm). Negative samples in the training dataset are set to those impressed products but not clicked by users, and posivtie samples are clicked.



\begin{table}[]

\caption{Statistics of the offline dataset. ( QL indicates Query Length, PL indicated Product Length. ) }
\label{stat}
\begin{tabular}{lcccc} \tiny\\

\toprule
Dataset            & \thead{Clothing,\\shoes\&jewelry} & Electronics & Toys\&Gift   & Fliggy      \\
\midrule
Users              & 36421                   & 34489       & 76234       & 2029216     \\
Products           & 22521                   & 18397       & 63323       & 920487      \\
Queries            & 1998                    & 1499        & 645         & 82409       \\
Avg.QL   & 8.23                    & 8.74        & 6.21        & 13.41       \\
Avg.PL & 75.36                   & 95.36       & 196.3       & 53.67       \\
\hline
\multicolumn{2}{l}{\#Query-User   Pairs}     &             &             &             \\
\hline
Train              & 456521                  & 273890      & 1493132     & 26379808    \\
Valid/Test         & 4006/4151               & 753/890     & 10210/10131 & 83129/89936\\
\bottomrule
\end{tabular}
\end{table}


\textbf{Amazon\footnote{http://jmcauley.ucsd.edu/data/amazon}}
As there are no large-scale public datasets in the personalized product
search area, we use Amazon dataset as our experiment corpus, consistent with existing approaches \cite{guo2019attentive,ai2019zero}. In our experiments, we adopted the five-core version provided, whereby the remaining users and
products have at least five reviews, respectively. Besides, we selected three categories with different sizes: Clothing, Shoes \& Jewelry, Toys \& Games, Electronics. These datasets both contain several categories so that users may have different interests.  
Following the strategy in References \cite{he2016ups}, we extracted the users’ product purchasing behaviors based on their reviews, i.e., the products they reviewed are the ones they purchased. Our model uses the previously purchased products in a
neighboring window size to model the short-term user interests. We further filtered the dataset to make sure each user has at least 20 purchased products (i.e., 20 reviews). 

Following the same paradigm used in \cite{guo2019attentive}, we construct queries for each purchased item
with category information. This strategy is based on the
finding that directed product search is users’ search for a producer’s name, a brand, or a set of terms describing product category. We partitioned each of the four datasets into three
sets: training, validation and testing sets. We first extracted the user-product pairs from users’ reviews, and then extracted the queries for these products, getting triplets. For each dataset, the last purchasing transaction of each user is held for the testing set, the second last for the validation set, and the rest for the training set. Moreover, we hid the reviews of the validation and testing sets in the training phase to simulate the real-world scenarios. 

Since Amazon datasets do not provide negative samples, we adopt a negative sampling method to evaluate it. In detail, for each positive user-query-item triplet, we generate negative products with the same category as query randomly to simulate a real search scenario. Then, the generated products, query and user form negative triplet samples. We generate ten negative samples for each positive triplet.

%


\subsection{Competitors}
To verify the effectiveness of the proposed method, we compare it with following methods:

\begin{itemize}[leftmargin=*]

    \item \textbf{AEM} \cite{ai2019zero}: The Attention-based Embedding Model (AEM)
constructs query-dependent user embeddings by attending to users’
historical purchased items with the query.


    \item \textbf{ZAM} \cite{ai2019zero}: The Zero Attention Model (ZAM) extends AEM with a
zero vector and conducts differentiated personalization by allowing
the query to attend to the zero vector.

    
    
    \item \textbf{TEM} \cite{bi2020transformer}: The Transformer-based Embedding Model (TEM) [6] is a state-of-the-art model that encodes query and historically purchased items with a transformer and does item generation based on the encoded query-user information.
    
    \item \textbf{DeepFM} \cite{guo2017deepfm}: It combines the explicit high-order interaction
module with deep MLP module and traditional FM module,
and requires no manual feature engineering.
    
    \item \textbf{DIN} \cite{zhou2018deep}: It utilizes attention mechanism to activate relevant users’ behaviors with respect to corresponding targets and learns an adaptive representation vector for users’ interests.
    
     \item \textbf{DIEN} \cite{zhou2019deep}: It adopts an interest extractor layer to capture temporal interests from users' historical behaviors and integrates GRUs with attention mechanism for further capturing the involved interests with respect to the target item.
     
     \item \textbf{ZAM+TDIE} : Compared with ZAM, we add a Target-dependent Interest Encoder (TDIE) to model the target-dependent user interest. The target-dependent user interest is concatenated with user representation to form a new user interest vector.
    
     \item \textbf{TEM+TDIE} : Do the same operation to TEM as ZAM+TDIE.
     \item \textbf{ALSTP} \cite{guo2019attentive}: It represent long and short-term user preferences with an attention mechanism applied to users’ recent interacted products and their global vectors. It is the sota long and short-term modeling method for personalized product search.

     
     
\end{itemize}


\subsection{Metrics}
In the experiments, we use AUC and logloss (or cross-entropy) \cite{shen2021sar,shen2022deep} to evaluate the ranking ability of all comparison models on the Fliggy dataset. The larger AUC and smaller Logloss mean better ranking performance. On public datasets, namely Amazon, the NDCG@10, MRR and MAP are used \cite{guo2019attentive}, since these datasets do not provide negative samples and the computation of AUC requires both positive and negative samples in a dataset.

\subsection{Implement details}
We use the Adam optimizer. Embedding size d is set as 16. Batch normalization is enabled for the MLP, and the activation function is ReLU. We use grid-search to find the best hyper-parameters. The optimal settings for our proposed implementation are: Batchsize is
512. Learning rate is 0.001. Feed forward networks (FFN) in IDM is set [128, 64, 32, 16], and FFN in short term interest fusing is set [128, 64, 32, 3]. $\lambda$ is 0.1 in both datasets. $Ts$ is 10 for Fliggy, 8 for Clothing,shoes\&jewelry, 14 for Electronics, 12 for Toys\&gift. $\beta$ is 0.55 for Fliggy, 0.35 for Clothing,shoes\&jewelry, 0.45 for Electronics, 0.6 for Toys\&gift.
Value of each experimental result is the average of 10 repeated tests.

\subsection{Main Comparison Results} \label{main_com_result}
We illustrate the overall performance on two adopted datasets in
Table \ref{main_comp}. From the results, we have the following observations:
\begin{itemize}[leftmargin=*]
 \item \textbf{It is important to extract the target-dependent user interest in product search.} AEM, ZAM, and TEM all model the correlation between query and user behaviors to represent user interests. However, the user interests are not always focused. For example, when a user issues a query "clothing", he does not express preferences for colors, brands, and styles. Users will click navy blue clothings instead of other colors in the result page, because the navy blue color is the most common among the products purchased by users in the history. If we only model the relationship between query and historical behaviors to extract user interest vectors, we will ignore this important point. It can be seen from the table that the methods of modeling query and user historical behavior such as AEM, ZAM, TEM and the methods of modeling target item and user historical behavior such as DIN and DIEN have different performances in different datasets. However, after adding the module for modeling the relationship between target item and user behavior sequence (TDIE), the performance is significantly improved, which shows the importance of modeling user interests from two perspectives at the same time.
 \item \textbf{Joint modeling of long and short-term interests does not always bring performance gains for product search.} 
 Although ALSTP is the SOTA long and short-term interests method in product search. However, there are many issues have not been addressed carefully. First, short-term interests play an important role for final prediction, but ALSTP only considers query-dependent interest, neglecting the target-dependent and causal-dependent interests. Second, simply concatenating long and short-term interests neglect the fact that long and short-term interests have varying degrees of importance in different situations. Third, modeling long and short-term interests entangled with each other will increase model redundancy and leads to inferior accuracy. Results demonstrate that ALSTP is not consistently effective across different metrics and datasets. For example, ALSTP is the best baseline on Clothing dataset, but its  NDCG performance is poorer than ZAM by about 5.35\% on Electronics. It indicates that without carefully address the above issues, modeling long and short-term interests can not get consistent improvements.


\item \textbf{HIFN can achieve
significant improvements over all datasets and metrics.} 
HIFN outperforms all competitors with
significant progress. Specifically, HIFN improves NDCG by about
0.07 on Clothing dataset and improves AUC by about 0.017
on Fliggy dataset, agaisnt SOTA methods. The consistent and
significant progress indicate that HIFN models long and short-term  user interests in a more proper manner than previous works.

\end{itemize}
\begin{table*}[]
\small
\caption{Overall performance on Fliggy and Amazon datasets.}
\label{main_comp}
\begin{tabular}{c|ccc|ccc|ccc|cc}
\toprule
Dataset & \multicolumn{3}{c|}{Clothing,shoes\&jewelry}                     & \multicolumn{3}{c|}{Electronics} & \multicolumn{3}{c|}{Toys\&Games} & \multicolumn{2}{c}{Fliggy} \\
\midrule 
Model   & MRR   & NDCG  & MAP  & MRR       & NDCG     & MAP      & MRR       & NDCG     & MAP      & AUC             & Logloss           \\ \midrule 
AEM      & 0.024 & 0.025 & 0.019                            & 0.269     & 0.291    & 0.265    & 0.068     & 0.072    & 0.062    & 0.735           & 0.598             \\
ZAM      & 0.025 & 0.027 & 0.021                            & 0.289     & 0.315    & 0.288    & 0.079     & 0.112    & 0.088    & 0.734           & 0.601             \\
TEM      & 0.027 & 0.026 & 0.023                            & 0.201     & 0.261    & 0.194    & 0.138     & 0.158    & 0.129    & 0.731           & 0.621             \\
DeepFM      & 0.020 & 0.028 & 0.016                            & 0.222     & 0.296    & 0.286    & 0.073     & 0.104    & 0.073    & 0.737           & 0.588             \\
DIN      & 0.022 & 0.029 & 0.018                            & 0.223     & 0.298    & 0.289    & 0.075     & 0.105    & 0.076    & 0.739           & 0.581             \\
DIEN     & 0.023 & 0.034 & 0.022                            & 0.288     & 0.319    & 0.301    & 0.099     & 0.127    & 0.118    & 0.741           & 0.579             \\
ZAM+TDIE & 0.031 & 0.036 & 0.026                            & 0.311     & 0.318    & 0.322    & 0.106     & 0.144    & 0.123    & 0.746           & 0.552             \\
TEM+TDIE & 0.032 & 0.034 & 0.028                            & 0.315     & 0.289    & 0.291    & 0.141     & 0.162    & 0.134    & 0.749           & 0.541             \\
ALSTP    & 0.034 & 0.037 & 0.029                            & 0.302     & 0.299    & 0.294    & 0.101     & 0.134    & 0.120     & 0.742           & 0.576             \\
HIFN     & \textbf{0.041} & \textbf{0.044} & \textbf{0.035}                            & \textbf{0.331}     & \textbf{0.342}    & \textbf{0.351}    & \textbf{0.162}     & \textbf{0.171}    & \textbf{0.143}    & \textbf{0.759}           & \textbf{0.533}      \\
\bottomrule 
\end{tabular}
\end{table*}


 
\begin{table}[]
\small
\caption{Study on Short-term Interests Extractor.}
\label{table2}
\begin{tabular}{c|ccc|cc}
\toprule
Dataset        & \multicolumn{3}{c|}{Clothing,shoes\&jewelry} & \multicolumn{2}{c}{Fliggy} \\
\midrule
Model          & MRR      & NDCG    & MAP     & AUC             & Logloss           \\
\midrule
HIFN$^{*}$           & 0.033    & 0.032   & 0.023   & 0.739            & 0.581             \\
HIFN$^{*}$+QDIE      & 0.037    & 0.038   & 0.028   & 0.746           & 0.553             \\
HIFN$^{*}$+TDIE      & 0.036    & 0.037   & 0.026   & 0.745           & 0.558             \\
HIFN$^{*}$+CDIE      & 0.033    & 0.032   & 0.024   & 0.744           & 0.561             \\
HIFN w/o CDIE & 0.039    & 0.040    & 0.032   & 0.753           & 0.541             \\
HIFN w/o TDIE & 0.036    & 0.039   & 0.031   & 0.751           & 0.549             \\
HIFN w/o QDIE & 0.037    & 0.038   & 0.028   & 0.749           & 0.552             \\
HIFN w/o gate & 0.034    & 0.035   & 0.024   & 0.743           & 0.563             \\
HIFN           & \textbf{0.041}    & \textbf{0.044}   & \textbf{0.035}   & \textbf{0.759}           & \textbf{0.533}    \\  
\bottomrule 
\end{tabular}
\end{table}

 \subsection{Study on Short-term Interests Extractor}
In this part, we will discuss how each module of short-term interests extractor impact the effectiveness of HIFN. The base model is HIFN$^{*}$, which using a mean pooling operator (No Attention) to aggregate the output of GRU layer. In particular, we consider the following settings: 1) add QDIE, 2) add TIDE, 3) add CDIE, 4) add QDIE and TIDE, i.e.,without CDIE, 5) add QDIE and CDIE, i.e., without TIDE, 6) add TIDE and CDIE, i.e, without QDIE. 7) replacing the fusing gate with mean pooling operator.

 \subsubsection{Ablation Study of each module}
In this section, we investigate the impact of Short-term Interests Encoder. As illustrated in Table \ref{table2} , base model performs the worst, showing that useful signals could be easily buried in noise without distilling. In addition, either QDIE, TDIE or CDIE can improve the AUC compared with the base model and the former has the most gain, while the latter has the least gain, demonstrating that capturing query-, target- and causal-dependent user interest all can bring gains and query-dependent user interest helps the most in prediction accuracy, which is also consistent with the characteristics of the product search. (Note that, adding CDIE module just have a weak improvement in Clothing but strong relatively in Fliggy. This is because Amazon only contains user purchase bahavior, while FLiggy contaions purchase and click, and CDIE is proposed to extract information from user's multiple historical hehavior types.)  4), 5) and 6) all achieve better performance than previous method, implying that the information extracted by these three interest extraction modules is complementary. In contrast, HIFN learns user short-term interest from three different perspective and fusing the output vector of them with a well-designed gate network, obtaining the highest gains. It shows that capturing query-, target- and causal-dependent user interest can improve the accuracy of prediction.  
\begin{table}[]
\small
\caption{Study on SIE with different query types.}
\label{query}
\begin{tabular}{c|c|c}
\toprule
Query Type    & Abstract & Precise \\
\midrule
Model         & AUC      & AUC     \\
\midrule
HIFN            & \textbf{0.748}    & \textbf{0.776}   \\
HIFN w/o QDIE & 0.745    & 0.759   \\
HIFN w/o TDIE & 0.739    & 0.772   \\
HIFN w/o CDIE & 0.742    & 0.768  \\
\bottomrule
\end{tabular}
\end{table}

 \subsubsection{Study on datasets of different query types}
In this section, we investigate the impact of different interest extractors in datasets of different query types. We first divide the Fliggy test set into two sets according to the type of query, one is the abstract search query, and the other is the precise (i.e. non-abstract) search query. We define the abstract search query as 1) destination,e.g., Hangzhou, Shanghai, Miami; 2) tag, e.g., playing, surfing. Different from precise search, users do not express obvious demand mentality when issuing the abstract query. As illustrated in Table \ref{query}, there are several points worth noting. On the one hand, HIFN w/o QDIE performs worst than 
the other two in precise query dataset, it implies that when user has clear requirements , extracting user interest with respect to query can obtain more gains. On the other hand, HIFN w/o TDIE performs worst than the other two in abstract query dataset, implying that when user has no clear intent, taget-dependent interest plays the most important role, and extracting user interest with respect to target item is an effective way to mining the intent from the multi-interest of users. It further validates that jointly modeling query-dependent, target-dependent, causal-dependent user interest can achieve significant performance.


\begin{table}[]
\small
\caption{Study on Long-term Interests Extractor.}
\label{table_long}
\begin{tabular}{c|ccc|cc}
\toprule
Dataset               & \multicolumn{3}{c|}{Clothing,shoes\&jewelry} & \multicolumn{2}{c}{Fliggy} \\
\midrule
Model                 & MRR      & NDCG    & MAP     & AUC             & Logloss           \\
\midrule
HIFN$^{\dagger}$                  & 0.035    & 0.036   & 0.029   & 0.745           & 0.554             \\
HIFN$^{\dagger}$+Target Attention & 0.036    & 0.037   & 0.03    & 0.746           & 0.553             \\
HIFN$^{\dagger}$+Query Attention  & 0.039    & 0.042   & 0.032   & 0.754           & 0.558             \\
HIFN$^{\dagger}$+User Attention   & /   & /    & /   & 0.752     & 0.564             \\
HIFN                  & \textbf{0.041}    & \textbf{0.044}   & \textbf{0.035}   & \textbf{0.759}           & \textbf{0.533}      \\
\bottomrule
\end{tabular}
\end{table}

\subsection{Study on Long-term Interests Extractor}
In this section, we investigate the impact of the Long-term Interests Extractor in HIFN, the base model is HIFN$^{\dagger}$, which extracts user behaviors using mean pooling operator (No Attention). In particular, we consider the following settings: 1) Target Attention: taking the target item as the query and the behavior item as the key in the attention mechanism; 2) Query Attention: the same long-term attention mechanism with the SOTA long and short-term methods in product search, ALSTP; 3) User Attention: take the user field feature as the query and the behavior item as the key in the attention mechanism.  

As can be seen from table \ref{table_long}, each attention mechanism can beat base model, implying that directly meaning pooling user long-term behaviors could not fully attend and exploit the useful information. The query attention mechanism used in ALSTP and user attention mechanism both get better performance than target attention, we suppose that Long-term interests of users are stable and change slowly, while user field and query field features represent user's intrinsic characteristic, which is consistent with the user's long-term interest to some extent. (Since there is no user profile feature in Amazon dataset, we do experiments about user attention only on Fliggy dataset). HIFN, leveraging short-term interests to activate long-term interests, achieves the best performance, we suspect that this is because short-term interests are a unified vector fusing user profile, query and recent behaviors, which can extract users' long-term interests more properly. 

\begin{table}[]
\small
\caption{Study on Interests Fusing Module.}
\label{table_IFM}
\begin{tabular}{c|ccc|cc}
\toprule
Dataset     & \multicolumn{3}{c|}{Clothing,shoes\&jewelry} & \multicolumn{2}{c}{Fliggy} \\
\midrule
Model       & MRR      & NDCG    & MAP     & AUC         & Logloss      \\
\midrule
concatenate & 0.035    & 0.036   & 0.029   & 0.750        & 0.539        \\
 $\alpha$=0.9         & 0.037    & 0.036   & 0.028   & 0.747       & 0.544        \\
$\alpha$=0.75        & 0.036    & 0.034   & 0.025   & 0.749       & 0.541        \\
$\alpha$=0.6         & 0.034    & 0.033   & 0.022   & 0.751       & 0.537        \\
$\alpha$=0.45        & 0.035    & 0.034   & 0.024   & 0.753       & 0.535        \\
$\alpha$=0.3         & 0.036    & 0.035   & 0.026   & 0.747       & 0.543        \\
$\alpha$=0.15        & 0.032    & 0.031   & 0.021   & 0.748       & 0.545        \\
HIFN        & \textbf{0.041}    & \textbf{0.044}   & \textbf{0.035}   & \textbf{0.759}       & \textbf{0.533}\\
\bottomrule
\end{tabular}
\end{table}

\subsection{Study on Disentanglement of Long and Short-Term Interests}
In order to prove the effectiveness of the disentanglement framework, we designed two kinds of experiments. Firstly, we will verify whether the framework can disentangle short-term and long-term interests well. Secondly, we will verify the impact of different proxy settings on the results.

\subsubsection{Counterfactual Evaluation}


If the importance of different factors change over time, learning disentangling representations of underlying factors is beneficial to final accuracy \cite{scholkopf2021toward}. In e-commerce product search, users will have various types of behaviors, such as purchase, favorite, click, the user preferences revealed by each type of behavior range from strong to weak, i.e purchase expresses strong preferences, while click expresses relatively weak preferences. Grbovic \cite{grbovic2018real} acknowledged that behaviors of higher costs tend to be more driven by users’ long-term interests, and behaviors of lower costs such as click indicate more about short-term interests. As we known, purchase (cost of money) have higher costs than favorite (cost of double clicks), while favorite have higher costs than click (only one click). Therefore, a set of experiments were designed to verify the effectiveness of the disentanglement. Because Amazon dataset only contains user purchase behavior, we verify the effect on Fliggy dataset. We firstly train a model with exposure-click dataset, and predicting on both clicks, favorites and purchases test dataset of Fliggy. Intuitively, it is expected that the weight for short-term interests is smaller in predicting purchase than predicting click when fusing the two aspects. We compare our proposed model HIFN with HIFN without self-supervised framework, and the SOTA long and short-term methods for product search, ALSTP.

Table \ref{table_idm} illustrates the AUC, Logloss and the average $\alpha$ for clicked items, favorited items and purchased items. We have the following findings.
First, HIFN w/o IDM achieves a better performance than
ALSTP, demonstrating that HIFN extracts and fuses long and short-term interests in a more proper and effective way than ALSTP. Second, although predicting purchase/favorite with model trained on click data is challenging, HIFN also gets a good results and beats other competitors in favorite test set and purchase test set.
Third, the average $\alpha$ of HIFN in click is larger than favorite, and favorite is larger than purchase, which is consistent with the previous argument, that behaviors of higher costs tend to be more driven by users’ long-term interest, and behaviors of lower costs such as click indicate more about short-term interest. However the average $\alpha$ of HIFN w/o IDM in favorite is samller than purchase and the average $\alpha$ of ALSTP in click is samller than purchase/favorite. It demonstrates that HIFN successfully disentangle the long and short-term interest to a certain degree, while HIFN w/o IDM and ALSTP entangles two aspects with each other.



\begin{table*}[]
\small
\caption{Study on Interest Disentanglement Module.}
\label{table_idm}
\begin{tabular}{c|ccc|ccc|ccc}
\toprule
Dataset      & \multicolumn{3}{|c|}{Click} & \multicolumn{3}{c|}{Favorite} & \multicolumn{3}{c}{Purchase} \\
\midrule
Model        & AUC    & Logloss  & AVG ( $\alpha$ )   & AUC     & Logloss   & AVG ( $\alpha$ )    & AUC     & Logloss   & AVG ( $\alpha$ )    \\
\midrule
HIFN         & \textbf{0.759}  & \textbf{0.533}    & 0.783 & \textbf{0.746}   & \textbf{0.569}     & 0.721  & \textbf{0.741}   & \textbf{0.573}     & 0.699  \\
HIFN w/o IDM & 0.753  & 0.549    & 0.637 & 0.741   & 0.578     & 0.526  & 0.737   & 0.586     & 0.672  \\
ALSTP        & 0.742  & 0.576    & 0.433 & 0.733   & 0.591     & 0.621  & 0.729   & 0.598     & 0.596 \\
\bottomrule
\end{tabular}
\end{table*}

\subsubsection{The effectiveness of long-term interests proxy updating}

Although Zheng \cite{zheng2022disentangling} proposes a simple strategy for proxy setting to disentangling user interests, we argue that this strategy ignores the nature that long-term interests will evolve slowly, so we propose an updating strategy for long-term interests proxy to address the issue. In this section, we study the effectiveness of the long-term interests proxy updating strategy on the four datasets with respect to different metrics. As shown in table \ref{table_proxy}, the performance with the long-term user preference updating strategy surpasses the one without updating. In detail, updating strategy improves NDCG by about 0.003, 0.002, and 0.003 in three Amazon datasets respectively, and improves AUC by about 0.007 in Fliggy dataset. This is in accordance with our assumption that the long-term user preference should update gradually.




\begin{table*}[]
\small
\caption{Study on Different Proxy Settings.}
\label{table_proxy}
\begin{tabular}{c|ccc|ccc|ccc|cc}
\toprule
Dataset          & \multicolumn{3}{c|}{Clothing} & \multicolumn{3}{c|}{Electronics} & \multicolumn{3}{c|}{Toys\&Games} & \multicolumn{2}{c}{Fliggy} \\
\midrule
Strategy            & MRR      & NDCG    & MAP     & MRR       & NDCG     & MAP      & MRR       & NDCG     & MAP      & AUC             & Logloss           \\
\midrule
without updating & 0.038    & 0.041   & 0.032   & 0.329     & 0.340     & 0.347    & 0.159     & 0.168    & 0.137    & 0.752           & 0.542             \\
with updating    & \textbf{0.041}    & \textbf{0.044}   & \textbf{0.035}   & \textbf{0.331}     & \textbf{0.342}    & \textbf{0.351}    & \textbf{0.162}     & \textbf{0.171}    & \textbf{0.143}    & \textbf{0.759}           & \textbf{0.533}  \\
\bottomrule 
\end{tabular}
\end{table*}

\subsection{Study on Interests Fusing Module}
Althogh ALSTP achieves the significant performance, we argue that concatenating long and short-term interests will lead to inferior performance. In HIFN, we propose to aggregate them adaptively. We compare HIFN with two different settings: 1) concatenation, 2) using a fixed $\alpha$ when combining the two aspects. The result is illustrated in Table \ref{table_IFM}
, we can discover that HIFN with IFM outperforms all different settings. These results verify the necessity of adaptive fusion of long and short-term interests.

\subsection{Online A/B Test}
 We also deploy HIFN on our platform for A/B test, where baseline model is TEM+TDIE (Previously, TEM+TDIE had demonstrated online that its results were superior to other competitors in Section \ref{main_com_result}). The proposed method achieves average $\boldsymbol{5.13\%}$ overall CTR gains over the baseline model in successive fifteen days, which is consistent with the offline evaluation results. Now, HIFN has been deployed online and serves the main traffic.

\section{Related Work}
\subsection{Product Search}
Aware of the importance of personalization in product search,
Ai et al. \cite{ai2017learning} proposed a hierarchical embedding model where they use a convex combination of the query and user vector to predict purchased items. Guo et al. \cite{guo2019attentive} represent long and short-term user preferences with an attention mechanism applied to users’ recent purchases and their global vectors. Recently, from the analysis of commercial search logs, Ai et al. \cite{ai2019zero} observed that personalization does not always have a positive effect. They further proposed a zero-attention model (ZAM) that can control the influence of personalization. However, the maximal effect personalization can have is equal to the query. Bi et al. \cite{bi2020transformer} found this limitation and proposed a transformer model to encode the query and historically purchased items where personalization can have none to full effect. 
These models construct user profiles by capturing dynamic user interests with respect to query. However, they ignore the target-dependent interest and causal-dependent interest, which will bring additional information for more accurate prediction.

\subsection{CTR Prediction}

Existing CTR prediction works mainly focus on single scenario modeling from the following aspects: 1) feature interaction (e.g., FM \cite{5694074}, DeepFM \cite{guo2017deepfm}); 2) user historical behavior (e.g., DIN\cite{zhou2018deep}, DIEN \cite{zhou2019deep}); Factorization Machine (FM) is proposed to model feature interactions explicitly, while previous generalized linear models such as Logistic Regression (LR) \cite{richardson2007predicting} and Follow-The-Regularized-Leader (FTRL) \cite{mcmahan2013ad} lack the ability to solve interaction issue. Wide\&Deep \cite{cheng2016wide} and DeepFM \cite{guo2017deepfm} combine low-order and high-order features to improve the performance. FmFM \cite{sun2021fm} makes each field feature have different embedding dimensions, so as to reduce the amount of model parameters and avoid over fitting problem. DIN \cite{zhou2018deep} utilizes the attention mechanism to capture relative interests from the user behavior sequence with regard to the candidate item. DIEN \cite{zhou2019deep} further uses a GRU structure to capture the evolution of user interest. Considering a single vector might be insufficient to capture complicated user patterns, DMIN \cite{xiao2020deep} models user's multiple interests by a special designed extractor layer.

\subsection{Long and Short-term Interests Modeling}
Long and short-term interests modeling has been widely studied in recommendation, several methods \cite{an2019neural,zhao2018plastic,hu2020graph} were proposed to explicitly differentiate between long and short-term interests. For example, Zhao \cite{zhao2018plastic} use matrix factorization for long-term interests and use RNN for short-term interests. Yu \cite{yu2019adaptive} develop a variant of LSTM for short-term interests and adopt asymmetric SVD for long-term interests. However, in product search, modeling long and short-interests become difficult because of the query user issued. ALSPT, which is the sota long and short-term interests modeling method in product search, represents long and short-term preferences with attention mechanisms applied to users’ recent behaviors and their global vectors, achieves a significant performance, but it still contains weakness about interests modeling and fusing.

\section{Conclusion}

In this paper, we introduce a novel CTR model named HIFN for product search, which consists of four elaborately designed modules including SIE, LIE, IFM and IDM . SIE aims to extract short-term interests from different aspects. LIE is proposed to extract long-term interests with respect to the output of SIE. Then IFM fuses the long and short-term interests with an adaptive manner. IDM is used to disentangle the two aspects of interests. These modules are integrated together in an end-to-end multi-task learning framework. Extensive offline and online experiments on real-world e-commerce platforms demonstrate the superiority of HIFN over SOTA methods.




\balance
\bibliographystyle{ACM-Reference-Format}
\bibliography{sample-base.bib}

\end{document}